# A Computational Approach to Evaluate the Effect of Shelter Construction Material and Fuel Load on the Fire Spread Behavior in Rohingya Refugee Camp

Md. Fahad Hossain Mishu[1], Rafia Rizwana Rahim[1], Md. Ashiqur Rahman[1,a]

[1]*Department of Mechanical Engineering, Bangladesh University of Engineering and Technology, Dhaka-1000, Bangladesh*

[a] Corresponding author: ashiqurrahman@me.buet.ac.bd

**Abstract.** Numerous fires break out, especially from January to March every year, destroying thousands of shelters in the Rohingya Refugee Camps. In this study, a computational approach has been taken to analyze the fire dynamic behavior of informal settlements (ISs) in Rohingya Refugee Camps. The design parameters for the mathematical model are determined based on a comprehensive survey carried out in Rohingya Refugee Camps in Cox's Bazar, Bangladesh. This survey is conducted on three camps having distinctive fuel features, i.e., Kutupalong Registered Camp (KRC), Camp 5, and Camp 4 Extension (Ex), in order to accumulate data on the type, density, arrangement of fuel, etc. Using the dataset of almost 500 shelters having 42 fuel characteristics features, simplified small-scale representation of each camp under fire had been modelled in Fire Dynamic Simulator (FDS). It is found that Camp 5 is more vulnerable to fire hazard than Camp 4 Ex and KRC because of the dense fuel arrangement, bamboo structure and negligible separation distance between ISs. KRC's brick exterior acts as a fire barrier, and thereby despite having the largest internal fuel load of all three camps, prevents further spreading of fire from the ignited enclosure. It is also found that type of building material and separation distance are crucial for fire spread and increasing fuel load density causes the ceiling temperature to rise higher and the flashover point to approach faster. The findings of our study can be helpful for informed decision making of fire safety measures in the Rohingya Refugee Camps.

## INTRODUCTION

Fire is a common scenario in informal settlements (ISs), e.g., slums, refugee camps etc. Fires in the Rohingya refugee camps are particularly severe due to overcrowding, poor infrastructure, and use of combustible materials for shelters (bamboo and tarpaulin) [1]. Additionally, the risk of fires is higher during the dry season between October and June [1]. The International Organization for Migration (IOM) in Bangladesh has been keeping track of the number of incidents that affect the Rohingya refugees in Cox's Bazar since 2019. According to them, in 2020, over 150 fires were reported in Rohingya Refugee Camp [2]. On 14 January 2021, a devastating fire broke out in Nayapara Registered Refugee Camp in Teknaf, Cox's Bazar. Over 3,492 persons lost their homes and belongings, as the fire completely gutted 600 shelters [3]. 'Save the Children' reported that there were over 150 fires reported in the camps in 2021, "a staggering 180% increase on the 84 fires seen in 2020" [4]. This list keeps increasing.

The average population density in Rohingya Refugee Camps is 25 $m^2$/person [5], which is much higher than that suggested by the UN of 45 $m^2$/person [6]. With fire events occurring more and more often in refugee camps [7], understanding these incidents becomes critical, both experimentally and numerically. Unfortunately, there is a limited amount of information available on fire spread modelling of ISs. There have been some notable studies taking the ISs of Cape Town as reference [8]–[10]. The full-scale single shelter burning experiment was first conducted by Walls et al. [8], concluding that flashover in these small dwellings can occur in as little as one minute, which is consistent with observations from professional firefighters in the field. Cicione et al. [9] experimentally investigated a full-scale corrugated steel sheeting clad and a timber clad IS experiment, and later compared the results numerically. They reported a critical separation distance of 3m, below which fire spread between dwellings will have a 6% chance to occur. Wang et al. [10] took the study further by constructing four different real-scale compartments and igniting in a large-scale fire calorimeter. Koker et al. [11] conducted a full-scale fire spread experiment of a mock 20 dwelling settlement. Their results highlighted the critical hazard posed by the proximity of neighboring dwellings (1–2 m), with wind playing a primary role in directing and driving the fire spread process. Koker's study was numerically modeled by Cicione et al. using B-risk [12] and FDS [13], developing a semi-probabilistic method to determine the fire spread rates. Therefore, although an elaborate approach has been taken to estimate the fire spread phenomenon in ISs of Cape Town, due to the lack of experimental resources and past fire inventory map, studying the fire behavior of Rohingya Refugee camp has become a great challenge. Adequate survey data acquisition, pre-processing and development of a fuel inventory map having fuel type, density, fuel array arrangement etc. may pave the way to model fire spread behavior using the existing fire modelling software (e.g., FDS), which is the primary objective of our study.

TABLE 1. An overview of selection criterion for camp of interest and summary of the data obtained.

| Camp Name | Population | No. of Blocks | Camp Selection Criteria | | | | Total No. of Survey Data | |
|---|---|---|---|---|---|---|---|---|
| | | | Year of Establishment | Structure Material | Geographic Parameters | | Questionnaire | Picture of shelter (x5) |
| | | | | | Landscape | Separation Distance ft (m) | | |
| Camp 4 Ex | 6,691 | 4 | 2019 | Bamboo, Tarpaulin, wood etc. | Mostly Flat | 7 (2.1) | 40 | 50 |
| Camp 5 | 24,434 | 5 | 2017 | Bamboo, Tarpaulin, wood etc. | Hilly area with steep slopes | No visible space | 100 | 180 |
| KRC | 16713 | 7 | 1991 | Brick wall or Steel Sheet | Mostly Flat | 5 (1.5) | 100 | 170 |

## METHODOLOGY

### A. Survey Data Collection and Pre-Processing

**Selection of camps of interest:** A comprehensive survey was conducted in July 2022, to prepare the fuel inventory map featuring dwelling structure, fuel load density, placement of fuel etc. In the 5-day long survey, three camps having contrasting fuel inventory have been examined. They are- Camp 4 Ex, Camp 5, and Kutupalong Registered Camp (KRC). These camps were selected based on the features discussed in Table 1. Being the newest of all the camps in Kutupalong, Camp 4 Ex followed a standard design (Figure 1(a)) and an average separation distance of 7 ft as shown in Figures 1(b). On the other hand, Camp 5, one of the largest camps with a population of almost 25 thousand, is densely populated, having no visible space between two dwellings (Figure 1(c)). The Rohingya community there are as equally impoverished as those living in Camp 4 Ex. The shelters of both these camps are made of cellulose-based material like wood and bamboo, but those of Camp 5 do not follow any standard design. With a sharp contrast, the shelters in KRC (shown in Figure 1(d)) follow a shaded structure consisting of six dwellings under one roof, separated by brick wall or corrugated steel sheet. KRC began informally in 1991, making it the oldest of all camps and its dwellers somewhat self-sufficient. A lot more permanent furniture e.g., wooden table/chair/sofa, bed, wardrobe, showcase was found here, resulting in a much higher fuel load than that of the other two camps.

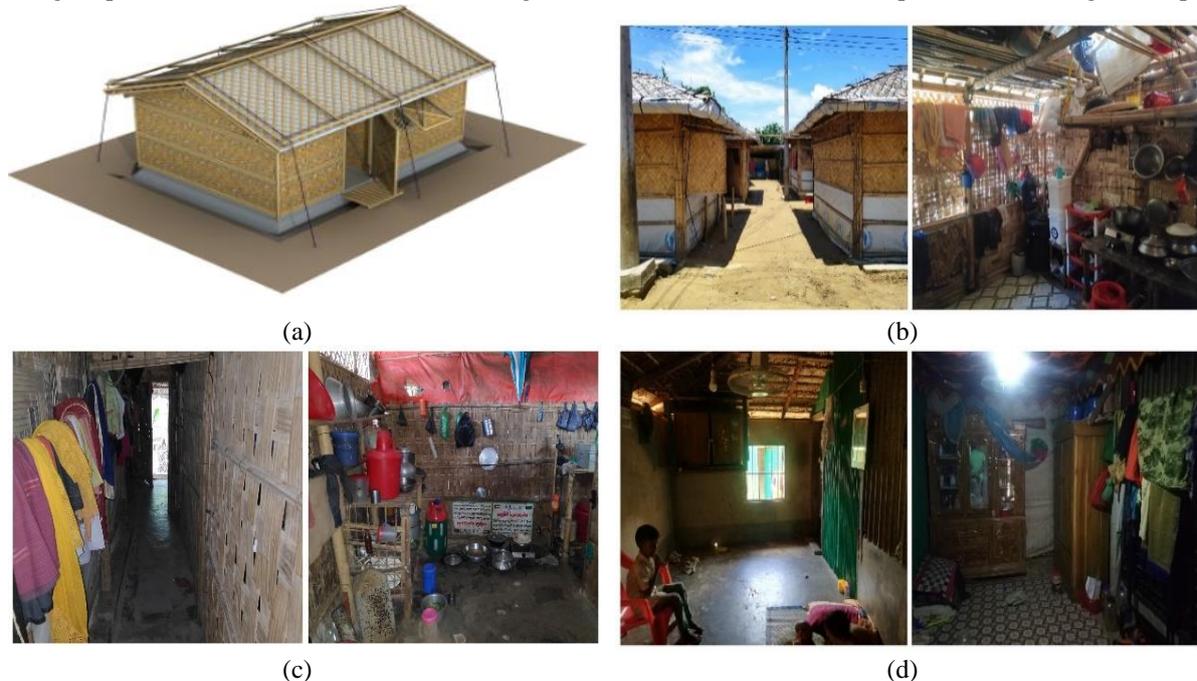

(a)                  (b)

(c)                  (d)

**FIGURE 1.** Shelter photos with interior and exterior fuel arrangement of (a) Standard Shelter Model [14], (b) Camp 4 Ex, (c) Camp 5, and (d) KRC [15]

**TABLE 2.** List of contributors of fuel load categorized according to survey data [21].

| Fuel Type | Contributor |
|---|---|
| General Plastics | chair, table, bucket, mug, melamine plate, polyethylene bag, water tank, battery, rack |
| PET plastics | water jug, water gallon, jar, sandal |
| Bamboo/Wood | baby swing, Borak bamboo, Muli bamboo, wooden furniture |
| Clothing | school bag, t-shirt, jeans, saree, lungi |
| Bedding | Mat, mosquito net, nylon rope, pillow, mattress, shack |
| Ignition source | Gas cylinder |

**Primary and Secondary Dataset – Questionnaire and Pictorial Data Collection and Their Tabulation:** During the survey, data was collected in two ways: a total of 240 handwritten questionnaires and 400 shelter photos. The data point distribution is described in Table 1. A handwritten form was created with 42 questions regarding the fuel arrangement, ignition source (i.e., gas cylinder placement), vegetation, shelter to shelter distance, human habits related to fire etc., and was filled up with the help of the local Rohingya community. For the pictorial data, shelters in each sub-block were randomly chosen and photos of both interior and exterior were taken. The main purpose of the pictorial data was to supplement the data from the filled-in questionnaires and to cross-check its relevancy. This dataset can be found in this zenodo repository [15].

At the end of the survey, both kinds of collected data were tabulated separately. Generally, the structure materials are provided by the government on a yearly basis and the exact amount of material has been described in [14]. We considered this as a benchmark to compare the actual fuel load associated with each camp. Pictorial data gave insight of fuel placement at the interior and the exterior of the shelter. It was found that for all 3 camps, 33 different types of interior objects are there to contribute to fuel load. These 33 objects were broadly classified into 6 categories- General plastics, Polyethylene terephthalate (PET) Plastics, Bamboo/Wood, Clothing, Bedding, and Ignition source, as shown in Table 2. Among the cellulose-based fuel load, Borak bamboo and Muli bamboo are widely used in Camp 4 Ex and Camp 5, where wood-based furniture is common in KRC. Clothing consisting of cotton, polyester, nylon, wool etc. was found and its amount varied with the age of the camp, KRC having the highest amount. A camp-wise distribution of different fuel and estimated amount of each contributor can be found in this zenodo repository [15].

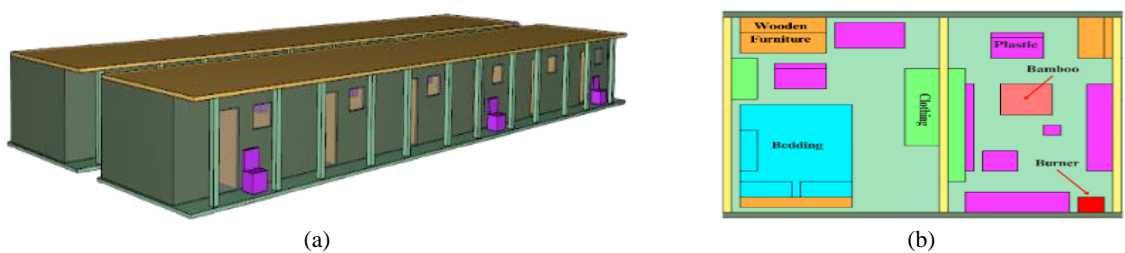

**FIGURE 2.** (a) Representative shelter model and (b) arrangement of internal fuel load of KRC.

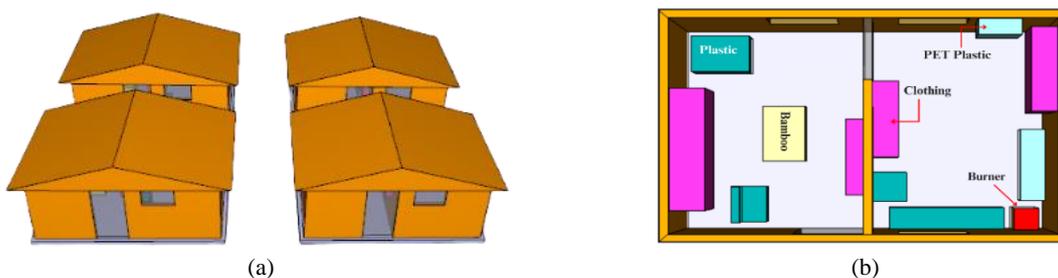

**FIGURE 3.** (a) Representative shelter model and (b) arrangement of internal fuel load of Camp 4 Ex & Camp 5.

**TABLE 3.** Dimensional parameters of each camp model for FDS simulation.

| Camp Models | Dwelling Design | Dwelling Dimensions (m³) | Separation Distance (ft, m) | Computational Domain Size (m³) | Cell Size (m³) | No. of Cells |
|---|---|---|---|---|---|---|
| Camp 4 Ex | Standard Design | 4.6×3×2.5 | 7, 2.1 | 12×10.5×3 | 0.001 | 358,440 |
| Camp 5 | Standard Design | 4.6×3×2.5 | 2, 0.6 | 11×9.5×3 | 0.001 | 303,490 |
| KRC | Shed Structure | 27×5×2.7 | 5, 1.5 | 28×12×3 | 0.001 | 985,320 |

## B. MATHEMATICAL MODELING

Based on the survey data, simplified model of the shelters in each of the three camps were developed using PyroSim, a graphical user interface (GUI) of FDS 6.7.8, where the tentative placement of fuels was decided based on the pictorial data collected of the interior of shelters. Figure 2(a) shows the simplified model of KRC, and its interior fuel arrangement is shown in Figure 2(b). Figure 3 is the representative shelter arrangement of both Camp 4 Ex and Camp 5, only differing in the separation distance between the shelters as stated in Table 1. Dimensional information of the developed models used to simulate fire spread in FDS can be found in Table 3.

Cell size sensitivity analysis was done to determine appropriate cell size for all the simulation cases. The 0.1D* criterion is widely used in fire dynamics [16], and thus an initial cell size 0.1D* is calculated as follows [17],

$$D^* = \left(\frac{\dot{Q}}{\rho_\infty C_p T_\infty \sqrt{g}}\right)^{\frac{2}{5}} \quad (1)$$

where D* is the characteristic fire diameter (m), $\dot{Q}$ is the heat release rate (HRR, kW), $\rho_\infty$ is the ambient air density (kg/m³), $T_\infty$ is ambient the temperature (K), $C_p$ is the specific heat [kJ/ (kg K)], and g is the gravitational acceleration [m/s²]. For an HRR per unit area (HRRPUA) of 1500 kW/m², a characteristic fire diameter of D* = 0.64 m is obtained, meaning a minimum cell size of 0.1D* = 0.064 m³ will provide appropriate result.

## RESULTS AND DISCUSSION

**Analysis of the Fuel Load in the Shelters:**

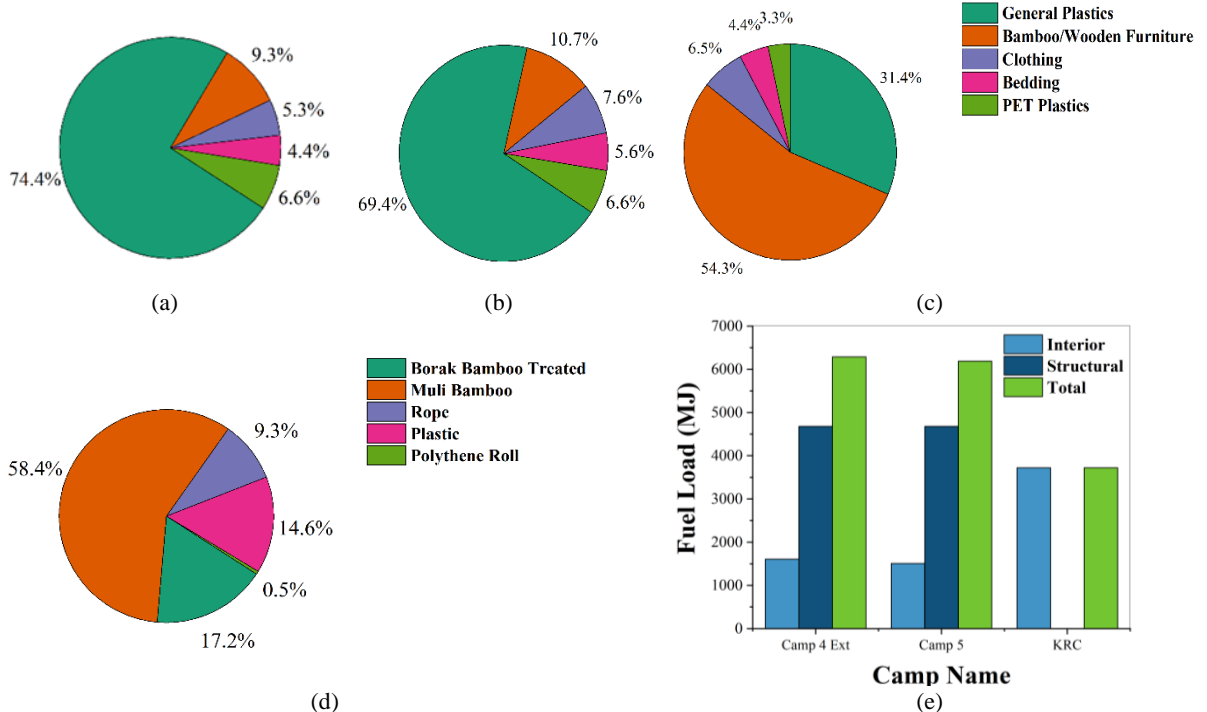

**FIGURE 4.** Percentage of each contributor in total fuel load of (a) Camp 4 Ex, (b) Camp 5, (c) KRC and (d) Standard Shelter Model; (d) Interior, exterior and total fuel load (MJ) for each camp of interest.

Four separate fuel load calculation for standard shelter, Camp 4 Ex, Camp 5, and KRC has been conducted, for which the SFPE handbook Table A.31 [18] was extensively used. Additionally, heat of combustion of treated Borak bamboo, Muli bamboo and Tarpaulin were taken from [19]–[21]. Mass of the contributing fuels was based on standard shelter model [14] and local market data. A detailed calculation of heat of combustion, mass value and fuel load calculation can be found in [15]. Figure 4 gives the overview of how each fuel type contributes to the fuel load of standard shelter and representative shelters of all three camps.

Camp 4 Ex and Camp 5 show almost similar distribution of fuel load (Figure 4(a-b)) where general plastic contributes the most in the fuel load. On the other hand, cellulose-based fuels, mostly wooden furniture, act as a primary fuel source for KRC (Figure 4(c)). The fuel load distribution of standard shelter in Figure 4(d) also differs from that of Camp 4 Ex and Camp 5. It appeared that apart from using bamboos as the main structural materials, dwellers of camp 4 and 5 also renovate their house at least 3-4 times a year, using mostly polypropylene rolls provided by the government [15]. This significantly changes the fuel load distribution dynamic of the camps built after the influx, making plastics to be the highest contributor to the fuel load. A comparison of fuel load of three camps has been shown in Figure. 4(e), where KRC is an outlier having no exterior fuel load. The main reason behind is that the permanent or semi-permanent structures of KRC are mostly constructed using brick or corrugated steel, which are non-combustible and hence do not contribute to the fuel load. On the other hand, bamboo, and tarpaulin-based exterior of camp 4 Ex and 5 makes them more vulnerable towards fire.

**Cell Sensitivity Analysis:** The cell size sensitivity analysis was performed using cell sizes of 0.05m, 0.064m and 0.1m, taking the results of the full-scale single shelter burning experiment [9]. From Figure 5(a), it can be seen that the experimental ceiling temperature vs. time curve is in very good agreement with the simulation when a cell size of 0.05m is used. Therefore, 0.05m cell size has been taken to simulate each camp model.

**Ceiling Temperature and Flashover of Ignited Shelter:** It has been found that the ceiling temperature of the ignited dwelling is significantly influenced by the structural fuel load and internal fuel density. The highly flammable bamboo exterior of shelters in Camp 4 Ex and Camp 5 increases the total fuel load which significantly affects the maximum ceiling temperature of its dwelling even though KRC dwellings have a higher internal fuel load (Figure 4(e)). Figure 5(b) shows that for Camp 5 and Camp 4 Ex, the ceiling temperature of the ignited dwelling is substantially higher than that of KRC. Within 4.5 minutes, the ceiling temperature of the ignited shelter of Camp 4 Ex hits its highest point of 1157°C. Camp 5 dwelling reaches its maximum temperature of 1135°C in roughly 5 minutes. On the other hand, the KRC dwelling's maximum ceiling temperature is much lower (548°C). However, because of its higher internal fuel load, it reached maximum temperature (within 3 minutes) quicklier than Camp 4 Ex and Camp 5. It takes camp 4 Ex and 5 dwellings to reach their flashover point at 2 and 2.6 minutes, respectively. The slightly lower ceiling temperature and higher flashover time of Camp 5 compared to Camp 4 Ex dwelling is due to the fact that Camp 5 shelters have lighter fuel distribution than Camp 4 Ex, as found from our survey. It should be noted that the classic definition for flashover (where flashover is defined as the rapid transition to full room involvement from the growth stage) criterion i.e., the hot gas layer temperature to reach 525°C, was used in this study [22].

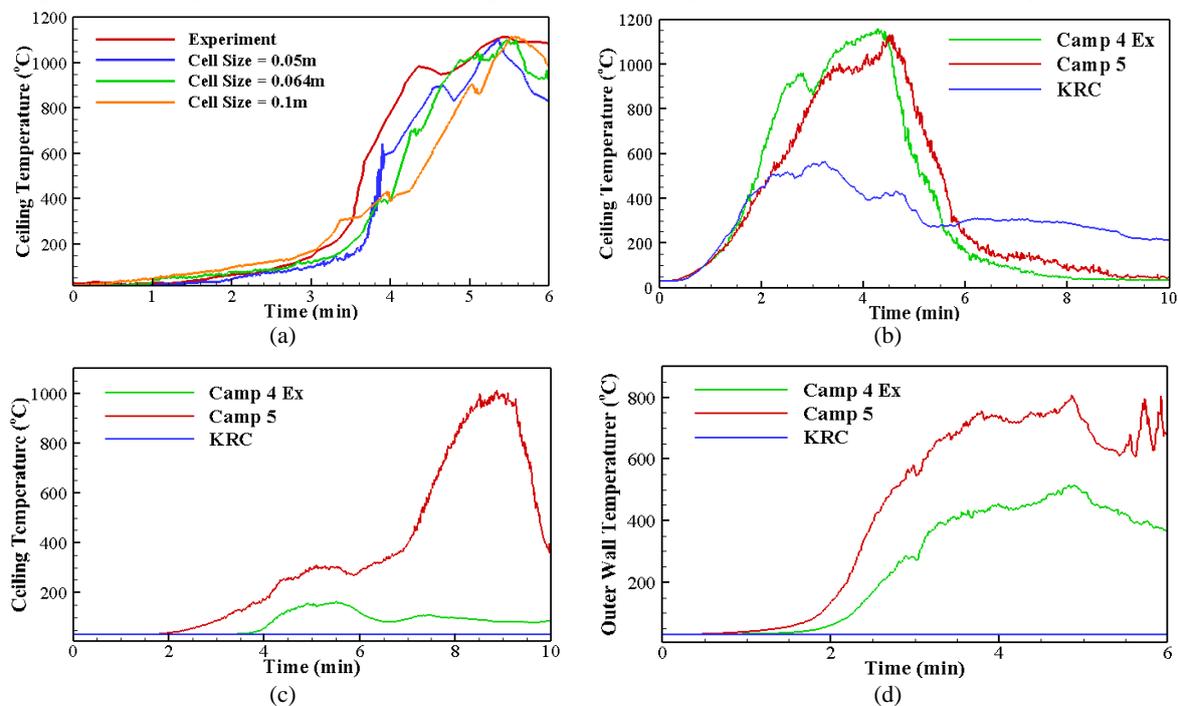

**FIGURE 5.** (a) Cell sensitivity analysis of the simulated domain with respect to single shelter experiment; (b) Variation of ceiling temperature with respect to time for ignited shelter; Variation of (c) outer wall temperature and (d) ceiling temperature of the nearest neighbour with respect to time for all three camps of interest.

**Wall and ceiling temperature of the nearest shelter:** The effect of construction material of the shelters is found to be more significant in fire spread to the nearest neighbor. Figure 5(c) illustrates that fire did not even propagate to the nearest dwelling in case of KRC and the outer wall temperature of the nearest dwelling remained constant. The brick walls and steel sheet cladding of KRC dwellings prevent fire from spreading to even the nearest wall, a sharp contrast from Camp 4 Ex and Camp 5. The outer wall of the nearest neighbor of the ignited shelter in Camp 4 Ex and Camp 5 catches fire within 1.5 and 1 minute from the start of the ignition, respectively. This is owing to the extremely short, only about 2 feet, separation distance of Camp 5. From Figure 5(d), a significant increase in the ceiling temperature (1022°C) of the nearest dwelling can be seen which reaches the flashover point after 7.5 minutes. Although the fire spreads to the nearby dwelling's wall in camp 4 Ex, the larger separation distance between the dwellings prevents it from reaching the flashover threshold. The peak ceiling temperature of adjacent dwelling in camp 4 extension is 163°C which is substantially lower compared to that of Camp 5. Therefore, it can be concluded that the construction material contributes considerably to the fuel load density of these shelters, and the separation distance and wall materials play a crucial role in the fire spread to the adjacent shelters. The bulk of Camps 5 dwellings are constructed using bamboo with a very close spacing between the shelters, making them more prone to fire.

## CONCLUSIONS

In this study, we developed a novel dataset on fuel inventory of Rohingya refugee camps in Coxs bazar, Bangladesh, containing a comprehensive analysis of the type, placement, and density of fuel for three camps having contrasting fuel inventory. Contributors of total fuel load have been classified into six broad categories and their distribution showed notable deviation from the fuel load characteristics of standard shelter model.

- From fire simulations using scaled-down models of shelters of three camps, we found that the shelter construction material plays a critical role in the fire spread as the bamboo-based shelter construction (as those in Camp 4 Ex and Camp 5) with a small separation distance between the shelters are observed to be the most vulnerable to fire spread.
- Dwellings with brick wall construction (such as those of KRC camp) slows down fire propagation to the nearest dwelling, eventually preventing it.
- It is also found that the combustibility of the shelter construction materials influences the fuel load density of the dwellings and results in a quicker flashover.

The findings of this study will be useful to conduct a more comprehensive fire simulations of the Rohingya refugee camps implementing other factors such as the effects of topography, wind, and vegetation, which can eventually lead to informed decision making for improving fire safety in the refugee camps.

## REFERENCES

[1] UNHCR, "UNHCR Bangladesh Operational Update, January 2022," Feb. 2022.
[2] BBC News, "Rohingya refugee camp fire: Several dead, hundreds missing and thousands homeless." Accessed: May 15, 2023. [Online]. Available: https://www.bbc.com/news/world-asia-56493708
[3] UNHCR, "Emergency response to the fire in Nayapara registered camp on 14 January 2021," Jan. 2021, Accessed: May 15, 2023. [Online]. Available: 6. https://reliefweb.int/report/bangladesh/emergency-response-fire-nayapara-registered-camp-14-january-2021
[4] Save the children, "One child killed and 1,000 children left homeless as fire rips through Rohingya refugee camp."
[5] UNHCR, "Camp profile: Rohingya refugee response Bangladesh," 2020.
[6] UNHCR Emergency Handbook, "Camp site planning minimum standards." Accessed: May 16, 2023. [Online]. Available: https://emergency.unhcr.org/emergency-assistance/shelter-camp-and-settlement/camps/camp-site-planning-minimum-standards
[7] Y. Kazerooni *et al.*, "Fires in refugee and displaced persons settlements: The current situation and opportunities to improve fire prevention and control," *Burns*, vol. 42, no. 5, pp. 1036–1046, Aug. 2016, doi: 10.1016/J.BURNS.2015.11.008.

[8] R. Walls, G. Olivier, and R. Eksteen, "Informal settlement fires in South Africa: Fire engineering overview and full-scale tests on 'shacks,'" *Fire Saf J*, vol. 91, pp. 997–1006, 2017, doi: https://doi.org/10.1016/j.firesaf.2017.03.061.

[9] A. Cicione, M. Beshir, R. S. Walls, and D. Rush, "Full-Scale Informal Settlement Dwelling Fire Experiments and Development of Numerical Models," *Fire Technol*, vol. 56, no. 2, pp. 639–672, 2020, doi: 10.1007/s10694-019-00894-w.

[10] Y. Wang, M. Beshir, A. Cicione, R. Hadden, M. Krajcovic, and D. Rush, "A full-scale experimental study on single dwelling burning behavior of informal settlement," *Fire Saf J*, vol. 120, p. 103076, Mar. 2021, doi: 10.1016/J.FIRESAF.2020.103076.

[11] N. de Koker *et al.*, "20 Dwelling Large-Scale Experiment of Fire Spread in Informal Settlements," *Fire Technol*, vol. 56, no. 4, pp. 1599–1620, 2020, doi: 10.1007/s10694-019-00945-2.

[12] A. Cicione, C. Wade, M. Spearpoint, L. Gibson, R. Walls, and D. Rush, "A preliminary investigation to develop a semi-probabilistic model of informal settlement fire spread using B-RISK," *Fire Saf J*, vol. 120, p. 103115, Mar. 2021, doi: 10.1016/J.FIRESAF.2020.103115.

[13] A. Cicione and R. Walls, *Towards a simplified fire dynamic simulator model to analyse fire spread between multiple informal settlement dwellings base on full-scale experiments*. 2019.

[14] "Shelter/NFI Sector (2021) Standard 10'x15' shelter for fire response in camps 8E, 8W, 9 Rohingya humanitarian crisis," Apr. 2021.

[15] R. R. Rahim and Md. F. H. Mishu, "Rohingya Refugee Camp Fuel Load Dataset (Camp 4Ex, 5, KRC)," Oct. 2023, doi: 10.5281/ZENODO.8401754.

[16] B. Merci and K. Van Maele, "Numerical simulations of full-scale enclosure fires in a small compartment with natural roof ventilation," *Fire Saf J*, vol. 43, no. 7, pp. 495–511, 2008, doi: https://doi.org/10.1016/j.firesaf.2007.12.003.

[17] K. B. McGrattan, R. McDermott, C. G. Weinschenk, K. J. Overholt, S. Hostikka, and J. E. Floyd, "Fire dynamics simulator technical reference guide volume 1 :: mathematical model," 2013. [Online]. Available: https://api.semanticscholar.org/CorpusID:126242428

[18] M. J. Hurley *et al.*, Eds., *SFPE Handbook of Fire Protection Engineering*, 5th ed. New York, NY: Springer, 2015. doi: https://doi.org/10.1007/978-1-4939-2565-0.

[19] A. O. Oyedun, T. Gebreegziabher, and C. W. Hui, "Mechanism and modelling of bamboo pyrolysis," *Fuel Processing Technology*, vol. 106, pp. 595–604, Feb. 2013, doi: 10.1016/J.FUPROC.2012.09.031.

[20] J. M. O. Scurlock, D. C. Dayton, and B. Hames, "Bamboo: an overlooked biomass resource?," *Biomass Bioenergy*, vol. 19, no. 4, pp. 229–244, Oct. 2000, doi: 10.1016/S0961-9534(00)00038-6.

[21] A. K. Panda, R. K. Singh, and D. K. Mishra, "Thermolysis of waste plastics to liquid fuel: A suitable method for plastic waste management and manufacture of value added products—A world prospective," *Renewable and Sustainable Energy Reviews*, vol. 14, no. 1, pp. 233–248, Jan. 2010, doi: 10.1016/J.RSER.2009.07.005.

[22] M. Beshir, M. Mohamed, S. Welch, and D. Rush, "Modelling the Effects of Boundary Walls on the Fire Dynamics of Informal Settlement Dwellings," *Fire Technol*, vol. 57, no. 4, pp. 1753–1781, 2021, doi: 10.1007/s10694-020-01086-7.